\documentclass[%
 reprint,
superscriptaddress,
 amsmath,amssymb,
 aps,
pra,
]{revtex4-2}

\usepackage{graphicx}% Include figure files
\usepackage{dcolumn}% Align table columns on decimal point
\usepackage{bm}% bold math
\usepackage[normalem]{ulem}
\usepackage{lineno}
\usepackage{soul}
\usepackage{lineno}
\usepackage{blindtext}
\usepackage{xcolor}

\begin{document}
%\linenumbers
\preprint{APS/123-QED}

% \title{Modulation instability and frequency comb generation in passive optical resonators with quadratic nonlinearity and spectral filter}
\title{Modulation instability and frequency-comb generation in hybrid quadratic-cubic resonators with spectral filtering}
\author{Minji Shi}
 \affiliation{
 Aston Institute of Photonic Technologies, Aston University, Birmingham B4 7ET, United Kingdom}
\author{Nicolas Englebert}
 \affiliation{
 Department of Electrical Engineering, California Institute of Technology, Pasadena, California 91125, USA}
\author{François Leo}
 \affiliation{
 OPERA-photonics, Université libre de Bruxelles (U.L.B.), Brussels, Belgium}
\author{Dmitry V. Skryabin}
 \affiliation{
 Department of Physics, University of Bath, Bath BA2 7AY, United Kingdom}
  \affiliation{
 Centre for Photonics, University of Bath, Bath BA2 7AY, United Kingdom}
 \affiliation{National Physical Laboratory, Teddington TW11 0LW, United Kingdom}
\author{Auro M. Perego}
 \email{a.perego1@aston.ac.uk}
 \affiliation{
 Aston Institute of Photonic Technologies, Aston University, Birmingham, B4 7ET, United Kingdom}

\date{\today}% It is always \today, today,

\begin{abstract}
We present an analytical and numerical investigation into the phenomenon of filter-induced modulation instability in passive hybrid optical resonators exhibiting quadratic and cubic nonlinearity. We show that asymmetric spectral losses, with respect to the continuous-wave solution frequency, can trigger sideband amplification in the normal dispersion regime. We calculate the parametric gain and demonstrate the associated pattern formation process. We furthermore show how this parametric process can be exploited to generate optical frequency combs with tunable repetition rate.
\end{abstract}

\maketitle

%\tableofcontents
\section{Introduction}
Optical frequency combs (OFCs) are optical rulers consisting of equally spaced coherent spectral lines. They find applications in precision metrology, optical clocks, spectroscopy, microscopy and telecommunications, among others \cite{book,Fortier}. OFCs can be generated through various platforms, including mode-locked lasers \cite{Cund,Diddams}, electro-optic modulators \cite{Parriaux:20}, and driven Kerr resonators \cite{Kip,PASQUAZI20181}. Quadratic nonlinearity resonators are a promising platform for OFC generation too \cite{Leo_2016,englebert2021parametrically,Mosca_2018,Zhang_2019,hong}.
In this paper, we propose a novel method for OFC generation in hybrid quadratic-cubic resonators by extending the concept of the gain-through-filtering (GTF) process originally proposed and demonstrated in Kerr cavities \cite{perego2021theory, Bessin_2019}. This method is based on the more general concept of filter-induced modulation instability (MI) obtained by the presence of asymmetric spectral losses for signal and idler waves \cite{gtl}. The GTF process facilitates the generation of sidebands through asymmetric spectral filtering relative to the pump. Frequency-asymmetric narrowband losses induce the creation of one sideband in proximity to the filter frequency, with the other sideband symmetrically positioned with respect to the pump. The cascaded generation of higher-order sidebands eventually leads to the formation of an OFC. By adjusting the detuning between the pump and filter frequencies, the OFC line spacing can be controlled. 

\section{Generalized parametrically driven nonlinear Schr\"odinger equation}
\begin{figure}[!h]
\centering
\includegraphics[width=0.35\textwidth]{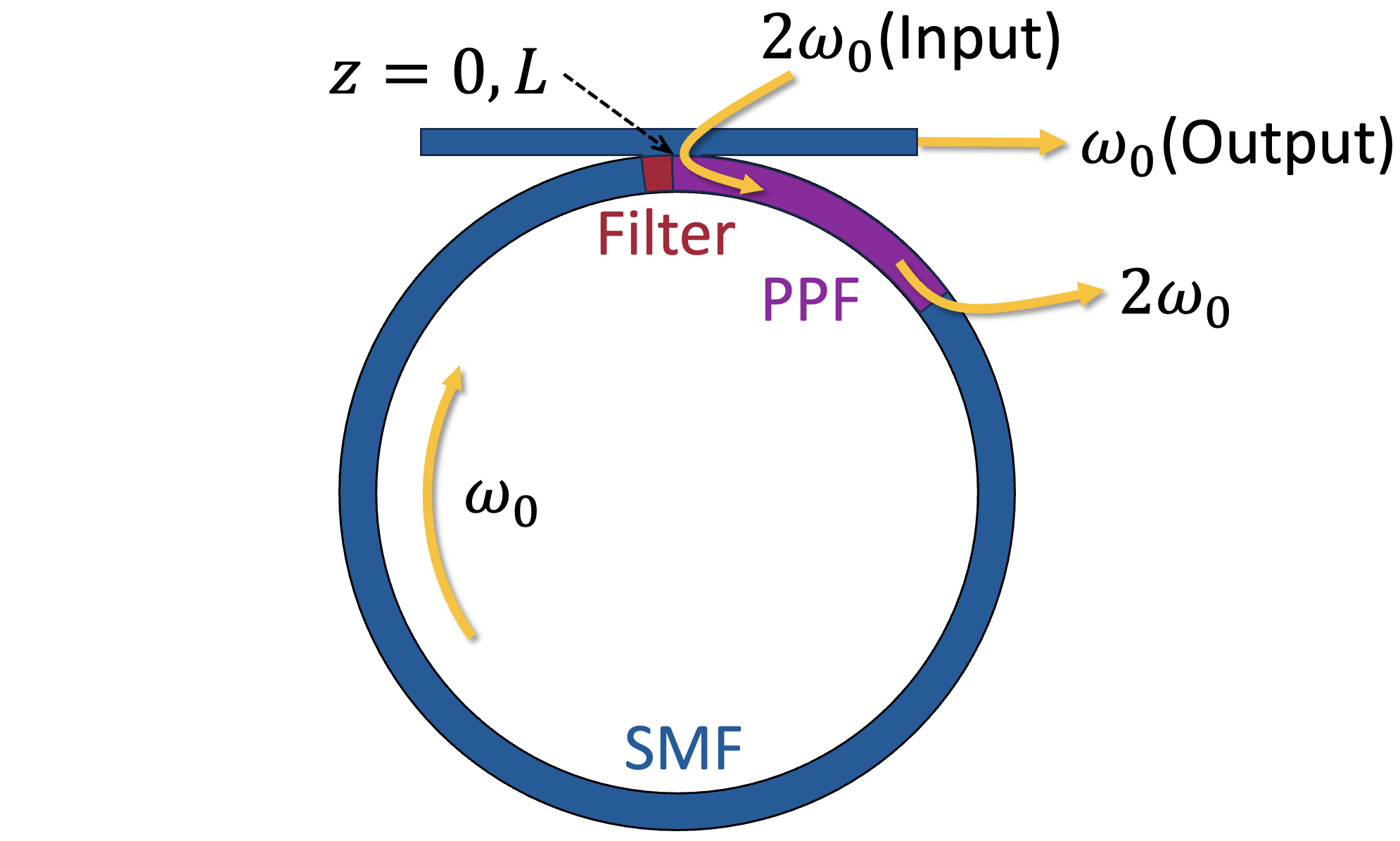}
\caption{Sketch of the setup consisting of single-mode fiber (SMF), a section of periodically poled fiber (PPF),  spectral filter, and input and output couplers.}\label{fig:setup}
\end{figure}
\begin{figure}[!h]
\centering
\includegraphics[width=0.48\textwidth]{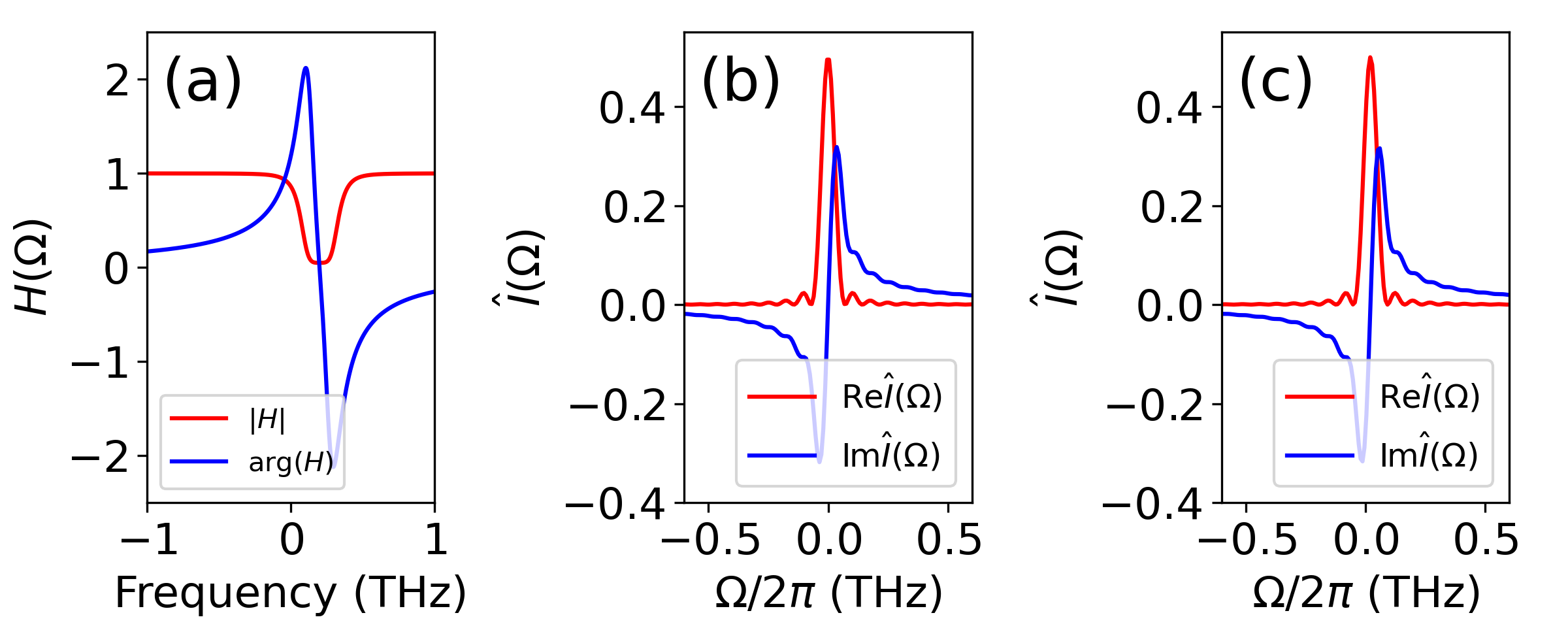}
\caption{(a) Amplitude and phase profile of $H(\Omega)$. (b) and (c) Profile of real and imaginary parts of $\hat I(\Omega)$ with $\Delta\beta=0$ in panel (b) and $\Delta\beta=46.5\text{ m}^{-1}$ in panel (c). Parameters used for the filter are $a = 0.6 \text{ rad/ps}$, $b=-3$, and $\Omega_f/(2\pi)=0.2 \text{ THz}$. Parameters used for the PPF are $\alpha^{(1)}=0.069\text{ km}^{-1}$ (corresponding to $0.3\text{ dB/km}$), $\Delta\beta_1=350\text{ ps/m}$, $\beta_{2,B}=10\text{ ps}^2\text{/km}$, and $L_1=4\text{ cm}$.}
\label{fig:filter_I}
\end{figure}

Quadratic resonators can be modeled either using coupled mode theory, which is especially relevant for integrated platforms \cite{Skryabin:20,skr1,skr2}, or by a mean-field parametrically driven nonlinear Schr\"odinger equation model which is frequently used for describing fiber cavities \cite{englebert2021parametrically,Leo_2016,Mosca_2018}. 
In this paper, we consider a ring cavity composed of a single-mode fiber (SMF), a section of periodically poled fiber (PPF) \cite{DeLucia:14, de2019single}, a spectral filter, the input and output couplers. A schematic representation of the setup is presented in Fig.~\ref{fig:setup}. The propagation of the optical wave envelope within the PPF and SMF is governed by the following equations respectively:
\begin{align}
\frac{\partial A_n}{\partial z} =&\left(-\frac{\alpha^{(1)}}2-i\frac{\beta_2^{(1)}}2\frac{\partial^2}{\partial t^2}\right)A_n\notag\\
&+i\kappa B_\text{in}A_n^*e^{-i\hat{k}(0) z}+i\kappa^2A_n^* \left[A_n^2\star J\right]\label{eq:PPLN}\\
\frac{\partial A_n}{\partial z} =& \left(-\frac{\alpha^{(2)}}2-i\frac{\beta_2^{(2)}}2\frac{\partial^2}{\partial t^2}+i\gamma^{(2)}|A_n|^2\right)A_n,
\end{align}
where
\begin{align}\label{eq:part_of_PPLN_eq}
& J(z,t)=\frac1{2\pi}\int_{-\infty}^{\infty}\frac{1-e^{-i\hat{k}(\Omega)z}}{\hat{k}(\Omega)}e^{-i\Omega t}\mathrm{d}\Omega,\\
& \hat{k}(\Omega)=-i\frac{\alpha^{(1)}}2+\Delta\beta-\Delta\beta_1\Omega-\frac{\beta_{2,B}}2\Omega^2.
\end{align}
$A_n(z,t)$ is the slowly varying envelope of the electric field with central angular frequency $\omega=\omega_0$, $t$ is the retarded time, $z$ is the spatial coordinate along the ring fiber cavity and $n$ is the roundtrip number.
$L_{1,2}$ are the length of PPF and SMF, respectively, and the total cavity length is denoted by $L=L_1+L_2$. Also, $\beta_2^{(1,2)}$ are group velocity dispersions at $\omega=\omega_0$, $\alpha^{(1,2)}$ are attenuation coefficients, $\kappa$ and $\gamma^{(2)}$ are the second- and third-order nonlinear parameters, respectively. In Eq.~(\ref{eq:PPLN}), $B_\text{in}$ is the driving wave amplitude of frequency $2\omega_0$ injected into the PPF with initial phase set to 0 without loss of generality; $\Delta\beta=2\beta(\omega_0)-\beta(2\omega_0)$ is the phase mismatch parameter; $\Delta\beta_1=\beta_1(2\omega_0)-\beta_1(\omega_0)$ is the difference of inverse group velocities at $\omega=2\omega_0$ and $\omega=\omega_0$; $\beta_{2,B}$ is the group velocity dispersion at driving frequency $2\omega_0$, whose contribution to phase-matching is negligible compared to the one given by $\Delta\beta_1$; and $\star$ denotes convolution with respect to time $t$, i.e., for two arbitrary functions $h(t)$ and $g(t)$, $h(t)\star g(t)=\int_{\infty}h(t-t')g(t')dt'$. 
$\Omega$ in Eq.~(\ref{eq:part_of_PPLN_eq}) is defined as the frequency relative to the driving frequency $2\omega_0$. As it appears solely in the integral, we have the flexibility to redefine it as desired. Henceforth, $\Omega$ will consistently denote the offset angular frequency with respect to $\omega_0$, i.e., $\Omega = \omega - \omega_0$. We note that the quadratic nonlinearity for the field at frequency $\omega=\omega_0$ acts as an effective Kerr term filtered by the convolution with $J$.
The filter at $z=z_F$ affects the wave amplitude in the following manner \cite{perego2021theory}:
\begin{equation}\label{eq:filter_time_domain_1}
A_n(z_F^+,t) = h(t)\star A_n(z_F^-,t),
\end{equation}
where $h(t)$ is the ﬁlter impulse response, whose frequency domain counterpart $H(\omega)=\int_{-\infty}^{\infty} h(t)e^{i\omega t}$ is called the filter transfer function. 
In the following, without loss of generality, a higher-order Lorentzian filter is considered whose transfer function reads
\begin{subequations}
\begin{align}
&H(\Omega) = e^{F(\Omega)+i\psi(\Omega)},\label{eq:H_exp}\\
&F(\Omega) = b\frac{a^4}{(\Omega-\Omega_f)^4+a^4},\\
&\psi(\Omega) = ba\frac{(\Omega-\Omega_f)[(\Omega-\Omega_f)^2+a^2]}{\sqrt2[(\Omega-\Omega_f)^4+a^4]},
\end{align}
\end{subequations}
where $a$ is related to the filter bandwidth (in rad$/$ps), $b < 0$ is a dimensionless parameter that governs the filter strength, specifically, the maximum attenuation, and $\Omega_f$ represents the filter center frequency. The profiles of the amplitude and the phase of $H(\Omega)$ are illustrated in Fig.~\ref{fig:filter_I}(a) for a specific set of parameters. In this context, in the vicinity of $z = z_f$, Eq.~(\ref{eq:filter_time_domain_1}) can be expressed as a differential equation as follows:
\begin{equation}
\frac{\partial A_n(z,t)}{\partial z} = \delta(z-z_F)(\Phi+i\Psi)\star A_n(z,t),
\end{equation}
where $\Phi$ and $\Psi$ are the inverse Fourier transform of $F(\Omega)$ and $\psi(\Omega)$ respectively.

The boundary condition for the envelope $A_n(z,t)$ can be expressed as
\begin{equation}
A_{n+1}(0, t) = \rho e^{i\varphi_0} A_n(L,t)+\theta e^{i\varphi_T} \sqrt{P_\text{in}},
\end{equation}
where $\rho^2$ and $\theta^2$ denote the power reflection and transmission coefficients of the coupler, respectively. Additionally, $P_\text{in}$ represents the power and $\varphi_T$ denotes the initial phase of the injected wave at frequency $\omega_0$, while $\varphi_0$ indicates the linear phase accumulated over one cavity round trip modulo $2\pi$. In this work, we consider the case where $P_\text{in}=0$. The inclusion of injection can be naturally considered for the investigation of more complex scenarios.

By defining $Z=z+nL$, $\alpha_T = -\ln\rho+[\alpha^{(1)}L_1+\alpha^{(2)}L_2)]/2$ and $\beta_2 = [\beta_2^{(1)}L_1+\beta_2^{(2)}L_2]/L$, $\gamma=\gamma^{(2)}L_2/L$, we finally obtain the generalized parametrically driven nonlinear Schr\"odinger equation with spectral filter governing the propagation of the optical wave envelope as follows:
\begin{align}\label{eq:lle}
L\frac{\partial A}{\partial Z}
=& (-\alpha_T+i\varphi_0) A + (\Phi+i\Psi)\star A \notag\\
&+ \left(-i\frac{\beta_2}2L\frac{\partial^2}{\partial t^2}+i\gamma L|A|^2\right)A\notag\\
& +i\kappa L_\text{eff}B_\text{in} A^* -(\kappa L_1)^2 \left(I\star A^2 \right)A^*,
\end{align}
%where $\theta'$ is a modified transmission coefficient, being a complex number, compensating the losses and the phase shift of the pump wave of frequency $\omega_0$, and $|\theta'|\approx 1$.
with
\begin{subequations}
\begin{align}
&I(t)=\int_0^{L_1}J(z,t)\mathrm{d}z=\frac1{2\pi}\int_{-\infty}^{\infty} \hat{I}(\Omega)e^{-i\Omega t}\mathrm{d}\Omega,\\
&L_\text{eff} = \int_0^{L_1} e^{-i\hat k(0)z'} dz' = L_1e^{-i\frac{\hat{k}(0)}2L_1}\text{sinc}\left[\frac{\hat{k}(0)L_1}2\right],
\end{align}
\end{subequations}
where $\hat I(\Omega)$ is defined as $\hat I(\Omega)=\frac{1-i\hat{k}(\Omega)L_1-e^{-i\hat{k}(\Omega)L_1}}{\hat{k}^2(\Omega)L_1^2}$ \cite{Leo_2016}. The profile of $\hat I(\Omega)$ is illustrated in Figs.~\ref{fig:filter_I}(b) and \ref{fig:filter_I}(c), corresponding to the cases of zero and nonzero $\Delta\beta$, respectively. We discuss further about $\hat I(\Omega)$ in Sec.~\ref{sec:modulation_instability}. We stress the fact that the driving of the cavity is performed by the $\chi^{(2)}$ response of the PPF --- forcing the field oscillations at frequency $\omega_0$ --- and that the PPF pump at frequency $2\omega_0$ is not resonant in the cavity.
%whose profile is shown in Fig.~\ref{fig:filter_I} b-c) for two sets of parameters. Note that $L_\text{eff}$ is real only if $\Delta \beta=0$.
%For the cavity in our consideration, $L_1, L_3 \ll L_2$, thus in the calculations, we keep $\beta_2=\beta^{(2)}_2$ and $\gamma=\gamma^{(2)}$ no matter what lengths of various devices are chosen.

\section{Stationary solutions}\label{sec:stationary_solutions}
\begin{figure}[!h]
\centering
\includegraphics[width=0.48\textwidth]{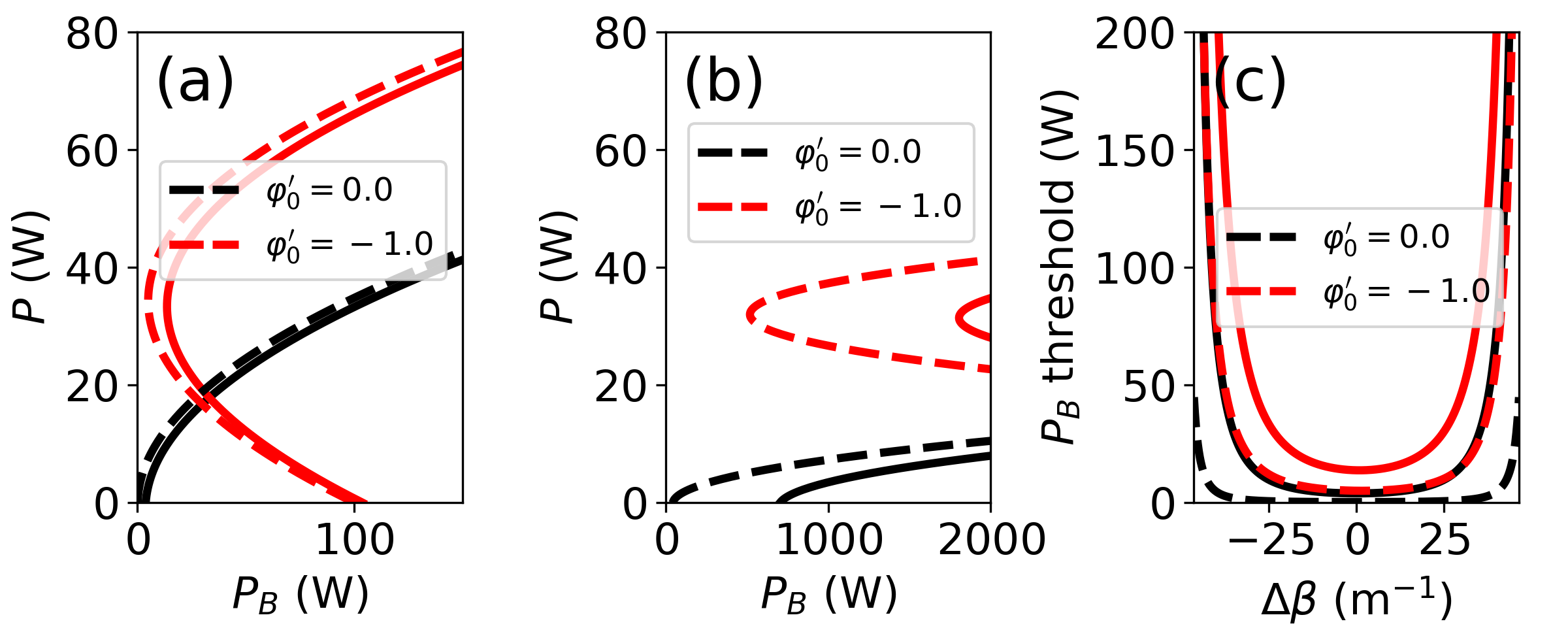}
\caption{(a) and (b) Power relationship between $P$ and $P_B$ of the nontrivial continuous waves with (solid lines) and without (dashed lines) filter considering $\Delta\beta=0$ in panel (a) and $\Delta\beta=46.5\text{ m}^{-1}$ in panel (b). (c) The minimum $P_B$ admitting positive $P$ for various $\Delta\beta$ values. Parameters used are $a = 0.6 \text{ rad/ps}$, $b=-3$, $\Omega_f/(2\pi)=0.2 \text{ THz}$, $\alpha^{(1)}=0.069\text{ km}^{-1}$ (corresponding to $0.3\text{ dB/km}$), $\Delta\beta_1=350\text{ ps/m}$, $\beta_{2,B}=10\text{ ps}^2\text{/km}$, $L_1=4\text{ cm}$, $\kappa=2.5 \text{ m}^{-1}\text{W}^{-1/2}$, $\gamma=1.4\text{ W}^{-1}\text{km}^{-1}$, $\alpha_T = 5\%$, and $L=20\text{ m}$.}
\label{fig:steady_power}
\end{figure}

If we assume that the continuous-wave (\textsc{cw}) solution of Eq.~(\ref{eq:lle}) is of the form $A(Z,t)=\sqrt{P}e^{i\xi}$, where $P$ and $\xi$ are the power and the phase of the wave, respectively, and substitute it into Eq.~(\ref{eq:lle}), we obtain a trivial solution $P=0$ along with nontrivial solutions ($P\ne 0$). The power and phase of the nontrivial solution can be obtained implicitly from the following two equations:
\begin{subequations}
\begin{align}
&\left|-\alpha_T+i\varphi_0+F(0)+i\psi(0)+P\left[i\gamma L-(\kappa L_1)^2\hat I(0)\right]\right|^2\notag\\
=&|\kappa L_\text{eff} B_\text{in}|^2,\\
&\frac{-\alpha_T+i\varphi_0+F(0)+i\psi(0)+P\left[i\gamma L-(\kappa L_1)^2\hat I(0)\right]}{-i\kappa L_\text{eff} B_\text{in}}\notag\\
=&e^{-i2\xi}.\label{eq:e_ixi}
\end{align}
\end{subequations}
The relationship between the intracavity power $P$ and the input power $P_B=|B_\text{in}|^2$ is illustrated in Figs.~\ref{fig:steady_power}(a) and \ref{fig:steady_power}(b) with solid lines, where $\varphi_0'=\varphi_0+\psi(0)$ is fixed to $0$ or $-1$. We observe that a nonzero $\Delta\beta$ results in a higher threshold of input power required to achieve a nontrivial \textsc{cw} solution. Figure~\ref{fig:steady_power}(c) further illustrates the relationship between the input power threshold and $\Delta\beta$. For reference, these figures also incorporate scenarios in which the filter is absent, represented by dashed lines.

\section{MI analysis}\label{sec:modulation_instability}
The study of the MI process of the intracavity \textsc{cw} solution is based on Eq.~(\ref{eq:lle}). The starting point is adding a perturbation term $\eta(Z,t)$ ($|\eta|^2\ll P$) to the stationary solution, i.e.,
\begin{align}\label{eq:pert_define}
A(Z,t) = \sqrt{P}e^{i\xi}+\eta(Z,t).
\end{align}
Substituting Eq.~(\ref{eq:pert_define}) into Eq.~(\ref{eq:lle}), and neglecting the higher-order terms of $\eta$ while retaining only the linear ones, we obtain a linear equation that characterizes the evolution of $\eta$ for both trivial and nontrivial continuous waves:
\begin{align}
L\frac{\partial \eta}{\partial Z} =& \left[-\alpha_T+i\varphi_0-i\frac{\beta_2}2L\frac{\partial^2}{\partial t^2} +i\gamma 2PL\right] \eta + (\Phi+i\Psi)\star\eta\notag\\
& +\left(i\gamma Pe^{2i\xi}L+i\kappa L_\text{eff} B_\text{in} -(\kappa L_1)^2\hat I(0)Pe^{2i\xi}\right)\eta^*\notag\\
& -(\kappa L_1)^2 2P\left(I\star \eta \right).\label{eq:pert_LLE}
\end{align}
In the Fourier domain, Eq.~(\ref{eq:pert_LLE}) becomes
\begin{align}
L\frac{\partial}{\partial Z}\begin{pmatrix}
\hat \eta(Z,\Omega)\\
\hat \eta^*(Z,-\Omega)
\end{pmatrix}=&M\begin{pmatrix}
\hat \eta(Z,\Omega)\\
\hat \eta^*(Z,-\Omega)
\end{pmatrix},
\end{align}
where
\begin{subequations}\label{eq:dsc_for_steady}
\begin{alignat}{2}
&M=\left[\begin{pmatrix}
d & 0\\
0 & d
\end{pmatrix}+\begin{pmatrix}
S & C\\
C^* & -S
\end{pmatrix}\right],\\
&d(\Omega) = -\alpha_T+F_\text{e}(\Omega)+i\psi_\text{o}(\Omega)-(\kappa L_1)^22P\hat I_+(\Omega),\\
&S(\Omega) = i\varphi_0+F_\text{o}(\Omega)+i\psi_\text{e}(\Omega)+i\frac{\beta_2}2L\Omega^2\notag\\
&\hspace{3.2em} +i2\gamma PL-(\kappa L_1)^22P\hat I_-(\Omega),\\
&C = i\gamma PLe^{i2\xi}+i\kappa L_\text{eff} B_\text{in} -(\kappa L_1)^2\hat I(0)Pe^{i2\xi},
% &\hspace{0.8em} = 
%\left[\alpha_T-i\varphi_0-F(0)-i\psi(0)\right]e^{i2\xi}\label{eq:C_expression},
\end{alignat}
\end{subequations}
and the subscripts ``e'' and ``o'' respectively indicate the even and odd part of the function $F(\Omega)$ and $\psi(\Omega)$, i.e., $F_{e,o}(\Omega)=[F(\Omega)\pm F(-\Omega)]/2$ and $\psi_{e,o}(\Omega)=[\psi(\Omega)\pm \psi(-\Omega)]/2$. $\hat I_\pm(\Omega)$ is defined as $\hat I_\pm(\Omega)=[\hat I(\Omega)\pm \hat I^*(-\Omega)]/2$. %, alternatively, $\hat I_+=\text{Re}\hat I_\text{e}+i\text{Im}\hat I_\text{o}$ and $\hat I_-=\text{Re}\hat I_\text{o}+i\text{Im}\hat I_\text{e}$. %Note that in Eq.~(\ref{eq:C_expression}), we have substituted Eq.~(\ref{eq:e_ixi}) into it.
The matrix $M$ has two eigenvalues $\lambda_\pm = d(Z,\Omega)\pm\sqrt{S(Z,\Omega)^2+|C(Z)|^2}$, and the MI power gain reads
\begin{align}\label{eq:gain_for_steady}
g(\Omega) =& \frac{2}{L}\left\{-\alpha_T+F_\text{e}(\Omega)- \text{Re}[\hat I_+(\Omega)](\kappa L_1)^22P\right.\notag\\
&\left.+\text{Re}(\sqrt{S(\Omega)^2+|C|^2})\right\}.
\end{align}
\begin{figure}[!h]
    \centering
    \includegraphics[width=0.48\textwidth]{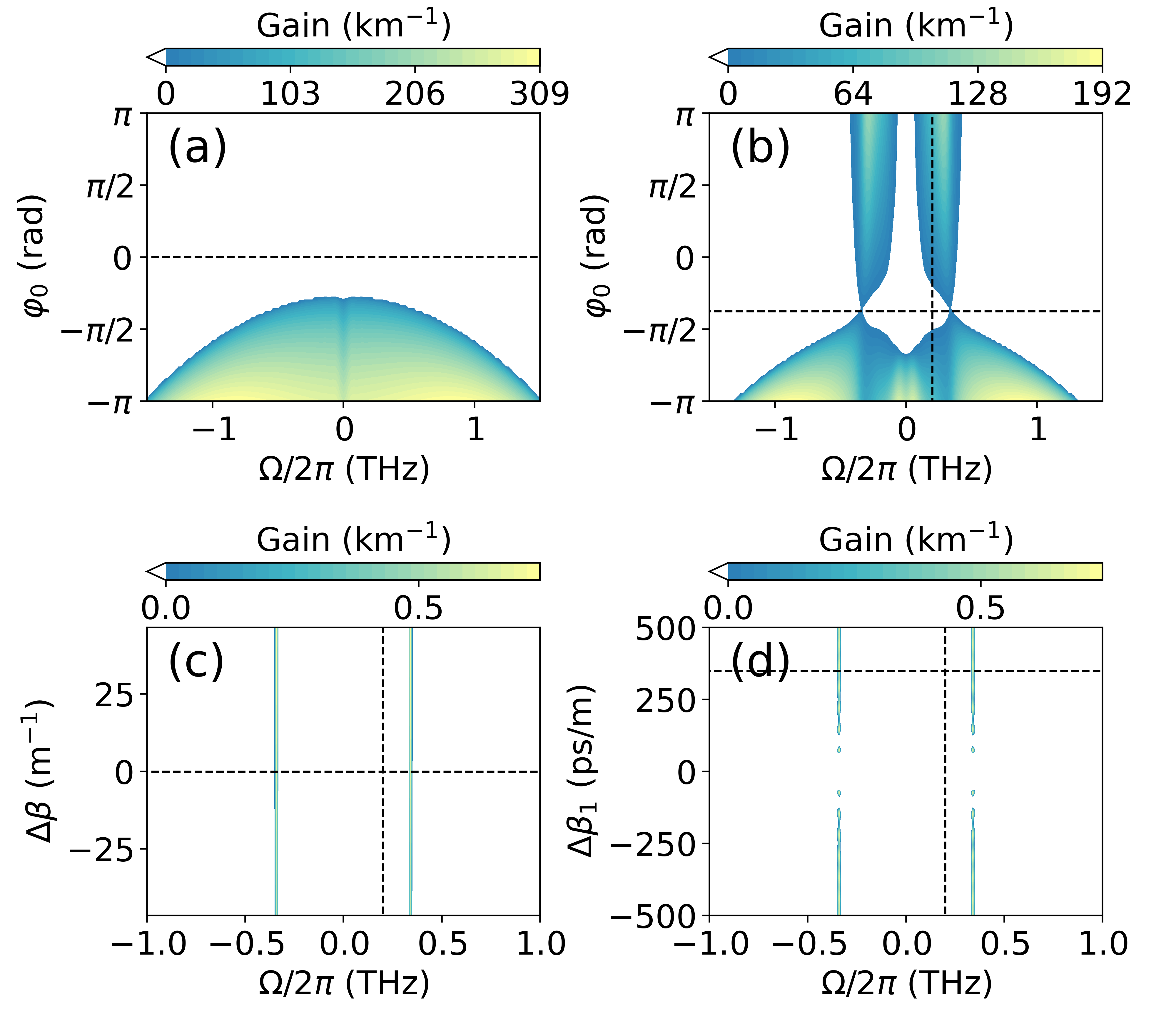}
     \caption{(a)-(b) MI gain map versus $\varphi_0$ in the absence (a) and presence (b) of a filter. (c)-(d) MI gain map versus $\Delta\beta$ (c) and $\Delta \beta_1$ (d) in the presence of a filter. The parameters used are $P=31\text{ W}$, $\beta_2=5\text{ ps}^2\text{km}^{-1}$, $\kappa=2.5 \text{ m}^{-1}\text{W}^{-1/2}$, $\gamma=1.4\text{ W}^{-1}\text{km}^{-1}$, $\alpha_T = 5\%$, $L=20\text{ m}$, $L_1=4 \text{ cm}$, $\alpha^{(1)}=0.069\text{ km}^{-1}$ (corresponding to $0.3\text{ dB/km}$), $\Delta\beta=0$, $\Delta\beta_1=350\text{ ps/m}$, $\beta_{2,B}=10\text{ ps}^2\text{/km}$, $a = 0.6 \text{ rad/ps}$, $b=-3$, $\Omega_f/(2\pi)=0.2 \text{ THz}$, and $\varphi_0=-\psi(0)$. The dashed horizontal lines indicate the parameters used for other panels, whereas the dashed vertical lines indicate the location of the filter center frequency.}\label{fig:with_and_without_filter}
\end{figure}From Figs.~\ref{fig:filter_I}(b) and \ref{fig:filter_I}(c), it can be observed that, for both $\Delta\beta=0$ and $\Delta\beta=46.5\text{ m}^{-1}$, $\text{Re}\hat I(\Omega)$ approximately behaves like an even function, whereas $\text{Im}\hat I(\Omega)$ behaves nearly as an odd function. As a result, $\text{Re}\hat I_\text{o}$ and $\text{Im}\hat I_\text{e}$ approach 0, so does $\hat I_-$.
Consequently, $\hat I$ influences the gain predominantly through $\text{Re}[\hat I_+(\Omega)]$, which leads to a significant reduction in gain around the central frequency, while not influencing the sidebands.

\begin{figure}[!h]
     \centering
     \includegraphics[width=0.48\textwidth]{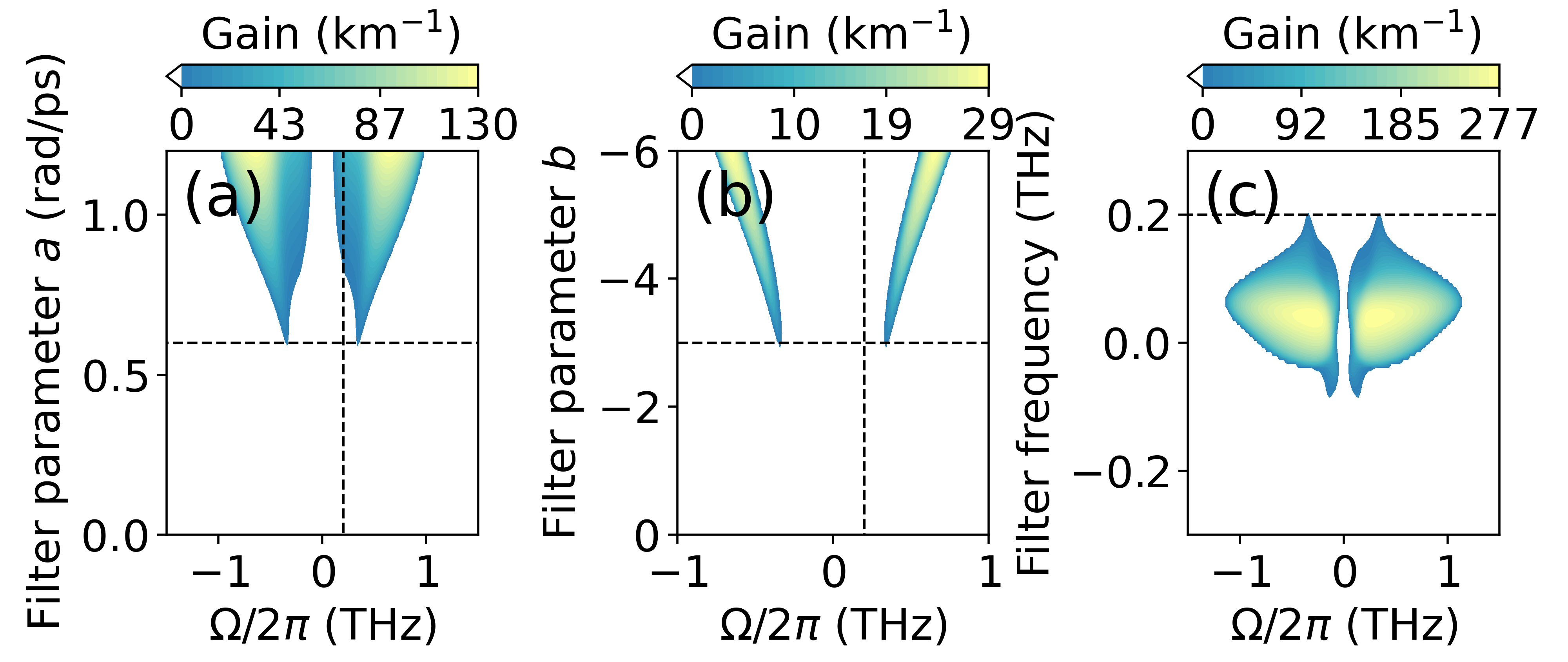}
     \caption{Analytical MI gain map in the presence of a filter in relation to various parameters. (a) MI gain versus filter width $a$. (b) MI gain versus filter strength $b$. (c) MI gain versus filter frequency $\omega_f/2\pi$. The parameters used are like those in Fig.~\ref{fig:with_and_without_filter}. The dashed horizontal lines indicate the parameters used for other panels, whereas the dashed vertical lines indicate the location of the filter center frequency.}
     \label{fig:gain_vs_paras1}
\end{figure}
\begin{figure}[!h]
     \centering
     \includegraphics[width=0.48\textwidth]{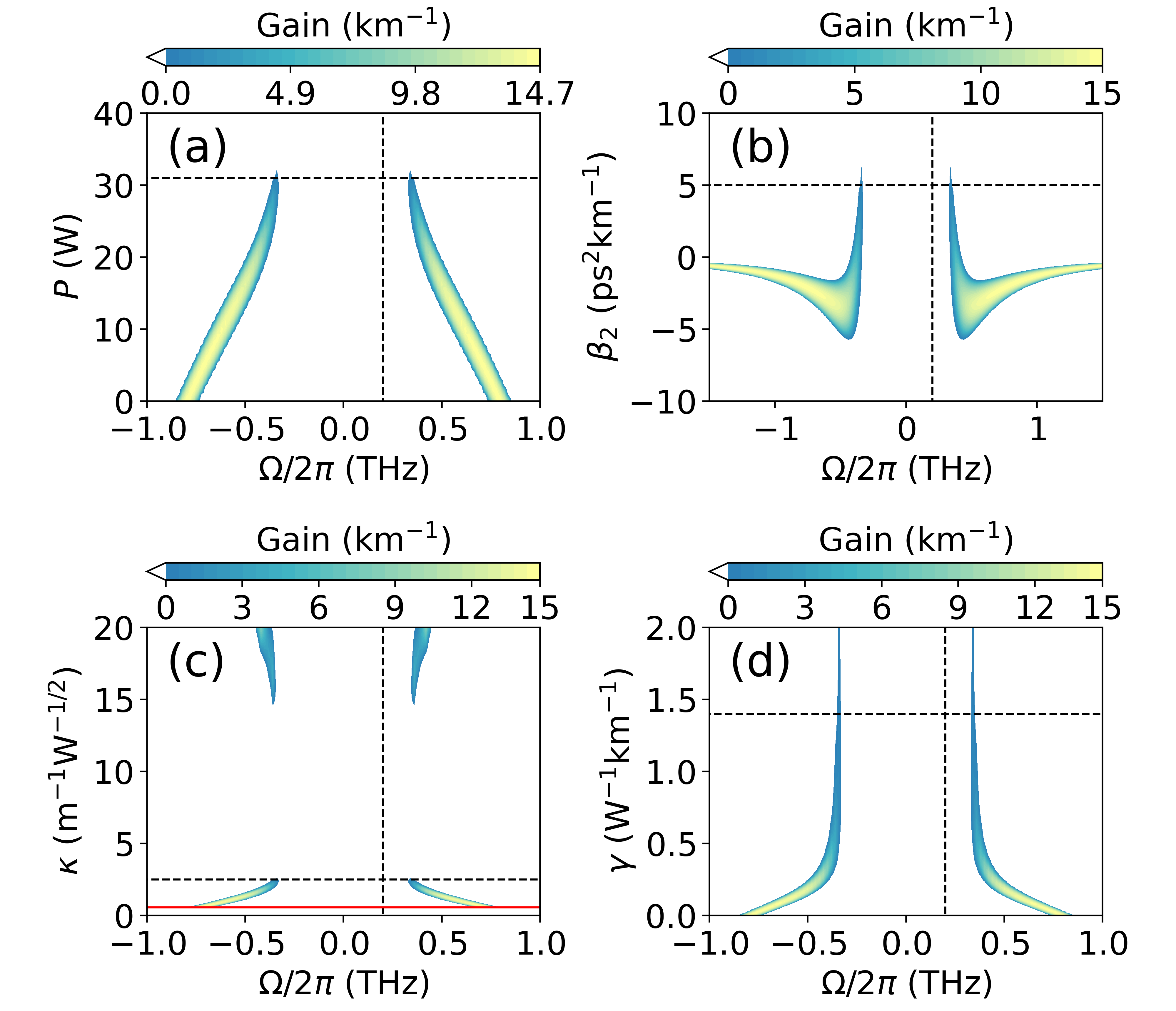}
 \caption{Analytical MI gain map in the presence of a filter in relation to various parameters. (a) MI gain versus intracavity power $P$. (b) MI gain versus $\beta_2$. (c) MI gain versus $\kappa$ with fixed input power $P_B=88$ W; the red solid line indicates the $\kappa$ threshold for achieving a \textsc{cw} solution.  (d) MI gain versus $\gamma$ with fixed $P_B=88$ W. The parameters used are like those in Fig.~\ref{fig:with_and_without_filter}. The dashed horizontal lines indicate the parameters used for other panels, whereas the dashed vertical lines indicate the location of the filter center frequency.}
     \label{fig:gain_vs_paras2}
\end{figure}

\section{MI gain and OFC generation for nontrivial \textsc{cw} solutions}
We first focus on the nontrivial \textsc{cw} solutions and study their MI gain and OFC generation.
To stress the impact of the filter, we present the analytical gain spectra as a function of $\varphi_0$ in Figs.~\ref{fig:with_and_without_filter}(a) and \ref{fig:with_and_without_filter}(b), exploring scenarios with and without the filter for the sake of comparison. It is evident that the filter enables MI for parameters' values where it would not be possible in the absence of the filter, around $\varphi_0'=0$, i.e., $\varphi_0=-\psi(0)$. The dependency of the MI gain on the phase mismatch parameter $\Delta\beta$ and the group velocity mismatch parameter $\Delta\beta_1$ is illustrated in Figs.~\ref{fig:with_and_without_filter}(c) and \ref{fig:with_and_without_filter}(d), respectively. We observe that $\Delta\beta$ does not have a significant impact, while a vanishing $\Delta\beta_1$ inhibits the MI gain.
% In the following, we only consider $\Delta\beta=0$, as the impact of $\Delta\beta$ on $\hat I(\Omega)$ is only to shift a bit the whole function along the frequency axis, as shown in Fig.~\ref{fig:filter_I} b-c). Note that a larger $\Delta\beta$ leads to a larger $P_B$, as shown in Fig.~\ref{fig:steady_power}, which is already too large for $\Delta\beta=46.5\text{ m}^{-1}$.
Moreover, Figs.~\ref{fig:gain_vs_paras1} and \ref{fig:gain_vs_paras2} illustrate the analytical gain maps as a function of additional parameters such as filter width, strength and position; dispersion, quadratic and Kerr nonlinearity; and intracavity power. Notably, the MI gain disappears and reappears as $\kappa$ increases, as illustrated in Fig.~\ref{fig:gain_vs_paras2}(b).
This phenomenon occurs due to the intracavity power going first below  the MI threshold (31 W) and then crossing the threshold again as $\kappa$ is increasing. Also, the gain sidebands induced by the filter are observed to persist even in the absence of fiber Kerr nonlinearity, as demonstrated in Fig.~\ref{fig:gain_vs_paras2}(d) in the limit $\gamma\rightarrow 0$, hence emphasizing the existence of MI in the limit where the PPF's quadratic nonlinearity dominates.
While in our study the pure Kerr term is due to the presence of a nonlinear fiber (SMF) in the resonator (see, e.g., Ref.~\cite{englebert2021parametrically}), we stress the fact that experimentally the dominant nonlinear term will depend on the specific setup considered.

\begin{figure}[!h]
     \centering
     \includegraphics[width=0.48\textwidth]{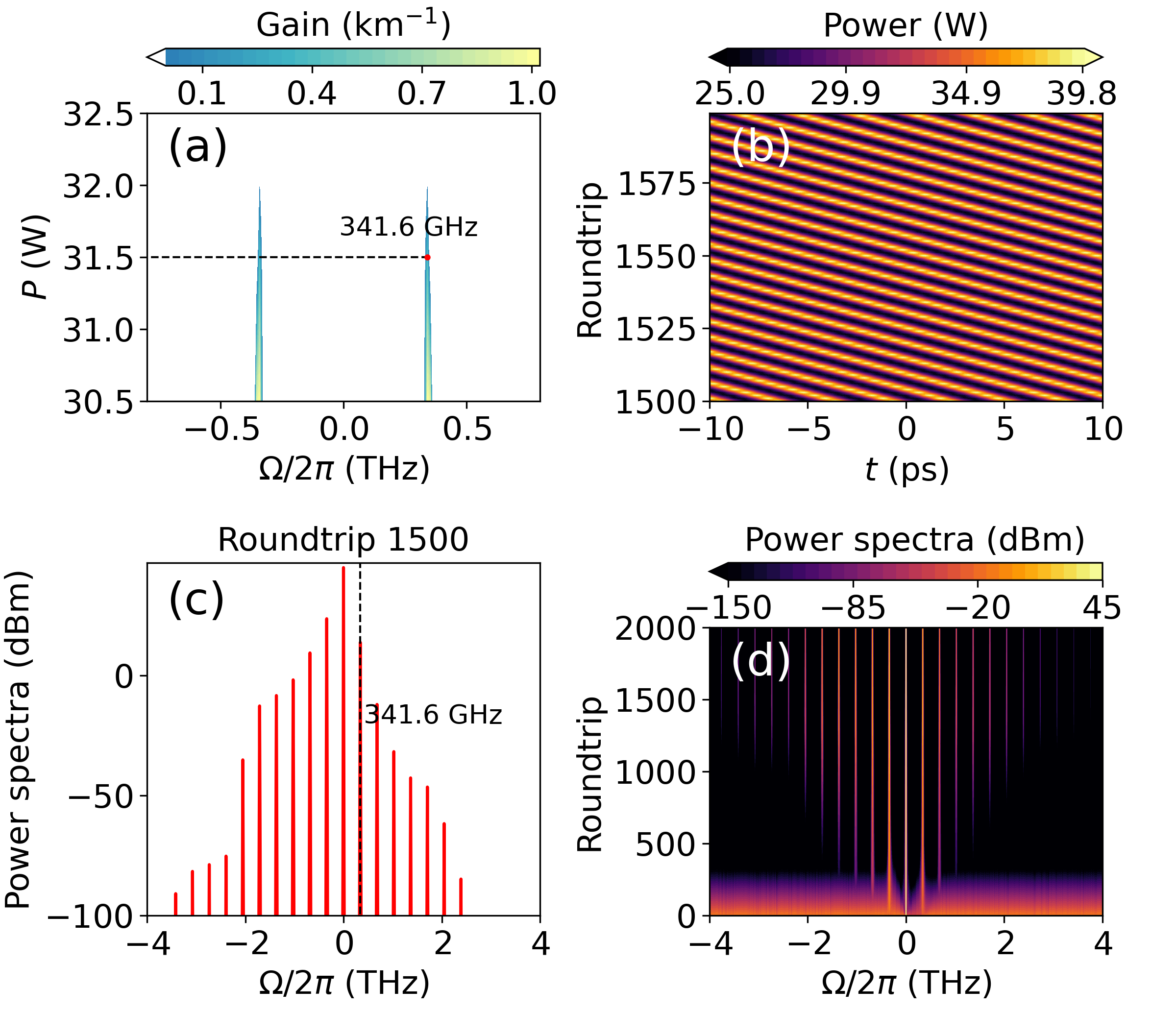}
     \caption{Numerical simulation results. (a) Analytical MI gain map in the presence of a filter versus the intracavity power $P$. The frequency possessing maximum gain at $P=31.5\text{ W}$ is marked. (b) The evolution of the field power between the $1500$th and $1600$th roundtrip. The simulation initial condition is the steady state to which a noise of power $1\text{ mW}$ has been added across the full spectrum. (c) The power spectrum at the $1500$th roundtrip. (d) The evolution of the power spectrum up to the $4000$th roundtrip. Other parameters used are the same as those in Fig.~\ref{fig:with_and_without_filter}.}
     \label{fig:comb_31}
\end{figure}
In order to validate the theory and to explore the nonlinear regime of the MI process we have performed numerical simulations by solving Eq.~(\ref{eq:lle}) using a standard split-step Fourier method. 
As it is possible to appreciate from Figs.~\ref{fig:comb_31}(a) and \ref{fig:comb_31}(c) the analytical estimation of the maximally unstable frequency agrees perfectly with numerical simulation results.
Furthermore, we observe that in the MI nonlinear stage a stable roll pattern can be formed [Fig.~\ref{fig:comb_31}(b)] whose spectral counterpart consists of an OFC whose line spacing is dictated by the frequency offset of the maximally unstable sideband with respect to the \textsc{cw} solution frequency, as shown in Figs.~\ref{fig:comb_31}(c) and \ref{fig:comb_31}(d).
As the MI gain position can be controlled by changing the detuning between $\omega_0$ and the filter frequency [Fig.~\ref{fig:gain_vs_paras1}(b)], this hints at the practical possibility of OFC generation with a tunable repetition rate. Roll pattern formation is achieved for parameters close to the MI gain threshold, and when the gain is very large, chaotic combs are generated in the nonlinear stage of MI.
%As shown in Fig.~\ref{fig:comb_31}, %for the parameters used in Fig.~\ref{fig:with_and_without_filter}, 
%we observe the frequency comb generated through the filter induced MI.

% As an example of small intracavity power, we show also the comb generation with $P=5\text{ W}$ in Fig.~\ref{fig:comb_5}.
% \begin{figure}[!h]
%      \centering
% \includegraphics[width=0.6\textwidth]{comb_5.0W_nround200.png}
%      \caption{a) analytical MI gain map in presence of filter versus $P$. The frequency possessing maximum gain at $P=5\text{ W}$ is marked; b) evolution of the field power up to $200$th roundtrip, the simulation initial condition is the combination of the steady state and a noise of power $1\text{ mW}$ across the full spectrum; c) power spectrum at $200$th roundtrip normalized to the power of the \textsc{cw} solution. d) evolution of the power spectrum normalised to the power of the \textsc{cw} solution up to $200$th roundtrip. The parameters used are the same as the default ones in Fig.~\ref{fig:with_and_without_filter}.}
%      \label{fig:comb_5}
% \end{figure}

\section{MI gain and OFC generation for trivial \textsc{cw} solutions}
\begin{figure}[!h]
    \centering
    \vspace{0.5em}
    \includegraphics[width=0.48\textwidth]{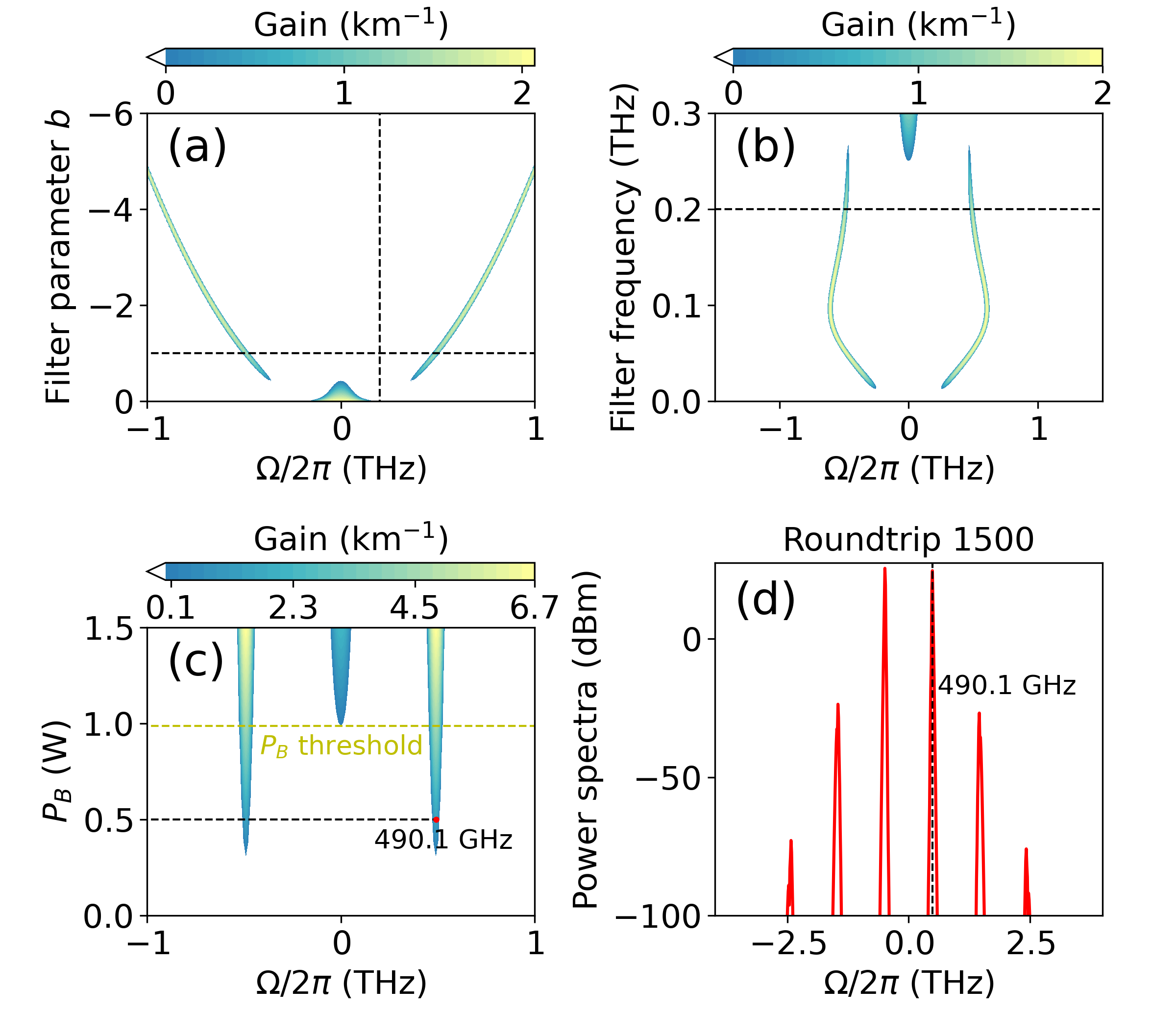}
     \caption{(a) and (b) MI gain map as a function of filter strength (a) and the filter center frequency (b). The dashed horizontal lines represent the parameters employed, whereas the vertical line in panel (a) indicates the position of the filter center frequency. (c) MI gain map versus the input power $P_B$. The frequency possessing maximum gain at $P_B=0.5\text{ W}$ is marked. The yellow dashed line indicates the power threshold for a positive nontrivial intracavity power $P$, which is also the stability threshold for the trivial solution. (d) The power spectrum at the $1500$th roundtrip. The parameters used are $P_B=0.5\text{ W}$, $\beta_2=5\text{ ps}^2\text{km}^{-1}$, $\kappa=2.5 \text{ m}^{-1}\text{W}^{-1/2}$, $\gamma=1.4\text{ W}^{-1}\text{km}^{-1}$, $\alpha_T = 5\%$, $L=20\text{ m}$, $L_1=4 \text{ cm}$, $\alpha^{(1)}=0.069\text{ km}^{-1}$ (corresponding to $0.3\text{ dB/km}$), $\Delta\beta=0$, $\Delta\beta_1=350\text{ ps/m}$, $\beta_{2,B}=10\text{ ps}^2\text{/km}$, $a = 0.6 \text{ rad/ps}$, $b=-1$, $\Omega_f/(2\pi)=0.2 \text{ THz}$, and $\varphi_0=-\psi(0)$.}\label{fig:fig8}
\end{figure}

As discussed in Sec.~\ref{sec:stationary_solutions}, there is a threshold of input power below which nontrivial solutions do not exist. Therefore, here we investigate the MI of trivial continuous waves below this threshold. As illustrated in Fig.~\ref{fig:fig8}(a), an increase in filter strength from 0 (corresponding to the introduction of the filter) results in the generation of new MI sidebands. From Fig.~\ref{fig:fig8}(b), we observe again the potential of OFC generation with a tunable repetition rate. The relationship between MI gain and the input power $P_B$ is illustrated in Fig.~\ref{fig:fig8}(c). We observe that the trivial solution is unstable with respect to zero-mode perturbations above the $P_B$ threshold (denoted by the yellow dashed line), and only nontrivial solutions are stable with respect to zero-mode perturbations above this threshold. At $P_B=0.5$ W, the maximum gain occurs at $\Omega/2\pi=490.1$ GHz, which agrees perfectly with the simulation results shown in Fig.~\ref{fig:fig8}(d), where a few-line OFC is generated.

\section{Conclusions}
We have presented analytical and numerical results regarding a filter-induced MI in hybrid quadratic-cubic optical resonators, enabled by asymmetric spectral losses for signal and idler waves. We have calculated analytically the parametric gain describing it as a function of several system parameters and obtained excellent agreement with numerical simulations. We have also shown how the GTF process can enable the generation of tunable optical frequency combs in quadratic-cubic resonators.

\section*{Acknowledgments}
AMP acknowledges support from the Royal Academy of Engineering through the Research Fellowship Scheme, and from EPSRC (project EP/W002868/1).
DVS acknowledges support from EPSRC (project EP/X040844)
 and the Royal Society (Grant No. IES/R3/223225).

\nocite{*}

\bibliography{apssamp}% Produces the bibliography via BibTeX.

\begin{thebibliography}{20}%
\makeatletter
\providecommand \@ifxundefined [1]{%
 \@ifx{#1\undefined}
}%
\providecommand \@ifnum [1]{%
 \ifnum #1\expandafter \@firstoftwo
 \else \expandafter \@secondoftwo
 \fi
}%
\providecommand \@ifx [1]{%
 \ifx #1\expandafter \@firstoftwo
 \else \expandafter \@secondoftwo
 \fi
}%
\providecommand \natexlab [1]{#1}%
\providecommand \enquote  [1]{``#1''}%
\providecommand \bibnamefont  [1]{#1}%
\providecommand \bibfnamefont [1]{#1}%
\providecommand \citenamefont [1]{#1}%
\providecommand \href@noop [0]{\@secondoftwo}%
\providecommand \href [0]{\begingroup \@sanitize@url \@href}%
\providecommand \@href[1]{\@@startlink{#1}\@@href}%
\providecommand \@@href[1]{\endgroup#1\@@endlink}%
\providecommand \@sanitize@url [0]{\catcode `\\12\catcode `\$12\catcode
  `\&12\catcode `\#12\catcode `\^12\catcode `\_12\catcode `\%12\relax}%
\providecommand \@@startlink[1]{}%
\providecommand \@@endlink[0]{}%
\providecommand \url  [0]{\begingroup\@sanitize@url \@url }%
\providecommand \@url [1]{\endgroup\@href {#1}{\urlprefix }}%
\providecommand \urlprefix  [0]{URL }%
\providecommand \Eprint [0]{\href }%
\providecommand \doibase [0]{https://doi.org/}%
\providecommand \selectlanguage [0]{\@gobble}%
\providecommand \bibinfo  [0]{\@secondoftwo}%
\providecommand \bibfield  [0]{\@secondoftwo}%
\providecommand \translation [1]{[#1]}%
\providecommand \BibitemOpen [0]{}%
\providecommand \bibitemStop [0]{}%
\providecommand \bibitemNoStop [0]{.\EOS\space}%
\providecommand \EOS [0]{\spacefactor3000\relax}%
\providecommand \BibitemShut  [1]{\csname bibitem#1\endcsname}%
\let\auto@bib@innerbib\@empty
%</preamble>
\bibitem [{\citenamefont {Perego}\ and\ \citenamefont {Ellis}(2024)}]{book}%
  \BibitemOpen
  \bibfield  {author} {\bibinfo {author} {\bibfnamefont {A.~M.}\ \bibnamefont
  {Perego}}\ and\ \bibinfo {author} {\bibfnamefont {A.}~\bibnamefont {Ellis}},\
  }\href@noop {} {\emph {\bibinfo {title} {Optical Frequency Combs, Trends in
  Sources and Applications}}}\ (\bibinfo  {publisher} {CRC, Boca Raton, FL},\
  \bibinfo {year} {2024})\BibitemShut {NoStop}%
\bibitem [{\citenamefont {Fortier}\ and\ \citenamefont
  {Baumann}(2019)}]{Fortier}%
  \BibitemOpen
  \bibfield  {author} {\bibinfo {author} {\bibfnamefont {T.}~\bibnamefont
  {Fortier}}\ and\ \bibinfo {author} {\bibfnamefont {E.}~\bibnamefont
  {Baumann}},\ }\bibfield  {title} {\bibinfo {title} {20 years of developments
  in optical frequency comb technology and applications},\ }\href@noop {}
  {\bibfield  {journal} {\bibinfo  {journal} {Commun. Phys.}\ }\textbf
  {\bibinfo {volume} {2}},\ \bibinfo {pages} {153} (\bibinfo {year}
  {2019})}\BibitemShut {NoStop}%
\bibitem [{\citenamefont {Cundiff}\ and\ \citenamefont {Ye}(2003)}]{Cund}%
  \BibitemOpen
  \bibfield  {author} {\bibinfo {author} {\bibfnamefont {S.~T.}\ \bibnamefont
  {Cundiff}}\ and\ \bibinfo {author} {\bibfnamefont {J.}~\bibnamefont {Ye}},\
  }\bibfield  {title} {\bibinfo {title} {Colloquium: Femtosecond optical
  frequency combs},\ }\href@noop {} {\bibfield  {journal} {\bibinfo  {journal}
  {Rev. Mod. Phys.}\ }\textbf {\bibinfo {volume} {75}},\ \bibinfo {pages} {325}
  (\bibinfo {year} {2003})}\BibitemShut {NoStop}%
\bibitem [{\citenamefont {Diddams}\ \emph {et~al.}(2020)\citenamefont
  {Diddams}, \citenamefont {Vahala},\ and\ \citenamefont {Udem}}]{Diddams}%
  \BibitemOpen
  \bibfield  {author} {\bibinfo {author} {\bibfnamefont {S.~A.}\ \bibnamefont
  {Diddams}}, \bibinfo {author} {\bibfnamefont {K.}~\bibnamefont {Vahala}},\
  and\ \bibinfo {author} {\bibfnamefont {T.}~\bibnamefont {Udem}},\ }\bibfield
  {title} {\bibinfo {title} {Optical frequency combs: Coherently uniting the
  electromagnetic spectrum},\ }\href@noop {} {\bibfield  {journal} {\bibinfo
  {journal} {Science}\ }\textbf {\bibinfo {volume} {369}},\ \bibinfo {pages}
  {eaay3676} (\bibinfo {year} {2020})}\BibitemShut {NoStop}%
\bibitem [{\citenamefont {Parriaux}\ \emph {et~al.}(2020)\citenamefont
  {Parriaux}, \citenamefont {Hammani},\ and\ \citenamefont
  {Millot}}]{Parriaux:20}%
  \BibitemOpen
  \bibfield  {author} {\bibinfo {author} {\bibfnamefont {A.}~\bibnamefont
  {Parriaux}}, \bibinfo {author} {\bibfnamefont {K.}~\bibnamefont {Hammani}},\
  and\ \bibinfo {author} {\bibfnamefont {G.}~\bibnamefont {Millot}},\
  }\bibfield  {title} {\bibinfo {title} {Electro-optic frequency combs},\
  }\href@noop {} {\bibfield  {journal} {\bibinfo  {journal} {Adv. Opt.
  Photon.}\ }\textbf {\bibinfo {volume} {12}},\ \bibinfo {pages} {223}
  (\bibinfo {year} {2020})}\BibitemShut {NoStop}%
\bibitem [{\citenamefont {Kippenberg}\ \emph {et~al.}(2011)\citenamefont
  {Kippenberg}, \citenamefont {Holzwarth},\ and\ \citenamefont
  {Diddams}}]{Kip}%
  \BibitemOpen
  \bibfield  {author} {\bibinfo {author} {\bibfnamefont {T.~J.}\ \bibnamefont
  {Kippenberg}}, \bibinfo {author} {\bibfnamefont {R.}~\bibnamefont
  {Holzwarth}},\ and\ \bibinfo {author} {\bibfnamefont {S.~A.}\ \bibnamefont
  {Diddams}},\ }\bibfield  {title} {\bibinfo {title} {Microresonator-based
  optical frequency combs},\ }\href@noop {} {\bibfield  {journal} {\bibinfo
  {journal} {Science}\ }\textbf {\bibinfo {volume} {332}},\ \bibinfo {pages}
  {555} (\bibinfo {year} {2011})}\BibitemShut {NoStop}%
\bibitem [{\citenamefont {Pasquazi}\ \emph {et~al.}(2018)\citenamefont
  {Pasquazi} \emph {et~al.}}]{PASQUAZI20181}%
  \BibitemOpen
  \bibfield  {author} {\bibinfo {author} {\bibfnamefont {A.}~\bibnamefont
  {Pasquazi}} \emph {et~al.},\ }\bibfield  {title} {\bibinfo {title}
  {Micro-combs: A novel generation of optical sources},\ }\href@noop {}
  {\bibfield  {journal} {\bibinfo  {journal} {Phys. Rep.}\ }\textbf {\bibinfo
  {volume} {729}},\ \bibinfo {pages} {1} (\bibinfo {year} {2018})}\BibitemShut
  {NoStop}%
\bibitem [{\citenamefont {Leo}\ \emph {et~al.}(2016)\citenamefont {Leo},
  \citenamefont {Hansson}, \citenamefont {Ricciardi}, \citenamefont {Rosa},
  \citenamefont {Coen}, \citenamefont {Wabnitz},\ and\ \citenamefont
  {Erkintalo}}]{Leo_2016}%
  \BibitemOpen
  \bibfield  {author} {\bibinfo {author} {\bibfnamefont {F.}~\bibnamefont
  {Leo}}, \bibinfo {author} {\bibfnamefont {T.}~\bibnamefont {Hansson}},
  \bibinfo {author} {\bibfnamefont {I.}~\bibnamefont {Ricciardi}}, \bibinfo
  {author} {\bibfnamefont {M.~D.}\ \bibnamefont {Rosa}}, \bibinfo {author}
  {\bibfnamefont {S.}~\bibnamefont {Coen}}, \bibinfo {author} {\bibfnamefont
  {S.}~\bibnamefont {Wabnitz}},\ and\ \bibinfo {author} {\bibfnamefont
  {M.}~\bibnamefont {Erkintalo}},\ }\bibfield  {title} {\bibinfo {title}
  {Walk-off-induced modulation instability, temporal pattern formation, and
  frequency comb generation in cavity-enhanced second-harmonic generation},\
  }\href@noop {} {\bibfield  {journal} {\bibinfo  {journal} {Phys. Rev. Lett.}\
  }\textbf {\bibinfo {volume} {116}} (\bibinfo {year} {2016})}\BibitemShut
  {NoStop}%
\bibitem [{\citenamefont {Englebert}\ \emph {et~al.}(2021)\citenamefont
  {Englebert}, \citenamefont {De~Lucia}, \citenamefont {Parra-Rivas},
  \citenamefont {Arab{\'\i}}, \citenamefont {Sazio}, \citenamefont {Gorza},\
  and\ \citenamefont {Leo}}]{englebert2021parametrically}%
  \BibitemOpen
  \bibfield  {author} {\bibinfo {author} {\bibfnamefont {N.}~\bibnamefont
  {Englebert}}, \bibinfo {author} {\bibfnamefont {F.}~\bibnamefont {De~Lucia}},
  \bibinfo {author} {\bibfnamefont {P.}~\bibnamefont {Parra-Rivas}}, \bibinfo
  {author} {\bibfnamefont {C.~M.}\ \bibnamefont {Arab{\'\i}}}, \bibinfo
  {author} {\bibfnamefont {P.-J.}\ \bibnamefont {Sazio}}, \bibinfo {author}
  {\bibfnamefont {S.-P.}\ \bibnamefont {Gorza}},\ and\ \bibinfo {author}
  {\bibfnamefont {F.}~\bibnamefont {Leo}},\ }\bibfield  {title} {\bibinfo
  {title} {Parametrically driven {K}err cavity solitons},\ }\href@noop {}
  {\bibfield  {journal} {\bibinfo  {journal} {Nat. Photonics}\ }\textbf
  {\bibinfo {volume} {15}},\ \bibinfo {pages} {857} (\bibinfo {year}
  {2021})}\BibitemShut {NoStop}%
\bibitem [{\citenamefont {Mosca}\ \emph {et~al.}(2018)\citenamefont {Mosca},
  \citenamefont {Parisi}, \citenamefont {Ricciardi}, \citenamefont {Leo},
  \citenamefont {Hansson}, \citenamefont {Erkintalo}, \citenamefont
  {Maddaloni}, \citenamefont {Natale}, \citenamefont {Wabnitz},\ and\
  \citenamefont {Rosa}}]{Mosca_2018}%
  \BibitemOpen
  \bibfield  {author} {\bibinfo {author} {\bibfnamefont {S.}~\bibnamefont
  {Mosca}}, \bibinfo {author} {\bibfnamefont {M.}~\bibnamefont {Parisi}},
  \bibinfo {author} {\bibfnamefont {I.}~\bibnamefont {Ricciardi}}, \bibinfo
  {author} {\bibfnamefont {F.}~\bibnamefont {Leo}}, \bibinfo {author}
  {\bibfnamefont {T.}~\bibnamefont {Hansson}}, \bibinfo {author} {\bibfnamefont
  {M.}~\bibnamefont {Erkintalo}}, \bibinfo {author} {\bibfnamefont
  {P.}~\bibnamefont {Maddaloni}}, \bibinfo {author} {\bibfnamefont {P.~D.}\
  \bibnamefont {Natale}}, \bibinfo {author} {\bibfnamefont {S.}~\bibnamefont
  {Wabnitz}},\ and\ \bibinfo {author} {\bibfnamefont {M.~D.}\ \bibnamefont
  {Rosa}},\ }\bibfield  {title} {\bibinfo {title} {Modulation instability
  induced frequency comb generation in a continuously pumped optical parametric
  oscillator},\ }\href@noop {} {\bibfield  {journal} {\bibinfo  {journal}
  {Phys. Rev. Lett.}\ }\textbf {\bibinfo {volume} {121}} (\bibinfo {year}
  {2018})}\BibitemShut {NoStop}%
\bibitem [{\citenamefont {Zhang}\ \emph {et~al.}(2019)\citenamefont {Zhang},
  \citenamefont {Buscaino}, \citenamefont {Wang}, \citenamefont {Shams-Ansari},
  \citenamefont {Reimer}, \citenamefont {Zhu}, \citenamefont {Kahn},\ and\
  \citenamefont {Lončar}}]{Zhang_2019}%
  \BibitemOpen
  \bibfield  {author} {\bibinfo {author} {\bibfnamefont {M.}~\bibnamefont
  {Zhang}}, \bibinfo {author} {\bibfnamefont {B.}~\bibnamefont {Buscaino}},
  \bibinfo {author} {\bibfnamefont {C.}~\bibnamefont {Wang}}, \bibinfo {author}
  {\bibfnamefont {A.}~\bibnamefont {Shams-Ansari}}, \bibinfo {author}
  {\bibfnamefont {C.}~\bibnamefont {Reimer}}, \bibinfo {author} {\bibfnamefont
  {R.}~\bibnamefont {Zhu}}, \bibinfo {author} {\bibfnamefont {J.~M.}\
  \bibnamefont {Kahn}},\ and\ \bibinfo {author} {\bibfnamefont
  {M.}~\bibnamefont {Lončar}},\ }\bibfield  {title} {\bibinfo {title}
  {Broadband electro-optic frequency comb generation in a lithium niobate
  microring resonator},\ }\href {https://doi.org/10.1038/s41586-019-1008-7}
  {\bibfield  {journal} {\bibinfo  {journal} {Nature}\ }\textbf {\bibinfo
  {volume} {568}},\ \bibinfo {pages} {373–377} (\bibinfo {year}
  {2019})}\BibitemShut {NoStop}%
\bibitem [{\citenamefont {Lu}\ \emph {et~al.}(2023)\citenamefont {Lu},
  \citenamefont {Puzyrev}, \citenamefont {Pankratov}, \citenamefont {Skryabin},
  \citenamefont {Yang}, \citenamefont {Gong}, \citenamefont {Surya},\ and\
  \citenamefont {Tang}}]{hong}%
  \BibitemOpen
  \bibfield  {author} {\bibinfo {author} {\bibfnamefont {J.}~\bibnamefont
  {Lu}}, \bibinfo {author} {\bibfnamefont {D.~N.}\ \bibnamefont {Puzyrev}},
  \bibinfo {author} {\bibfnamefont {V.~V.}\ \bibnamefont {Pankratov}}, \bibinfo
  {author} {\bibfnamefont {D.~V.}\ \bibnamefont {Skryabin}}, \bibinfo {author}
  {\bibfnamefont {F.}~\bibnamefont {Yang}}, \bibinfo {author} {\bibfnamefont
  {Z.}~\bibnamefont {Gong}}, \bibinfo {author} {\bibfnamefont {J.~B.}\
  \bibnamefont {Surya}},\ and\ \bibinfo {author} {\bibfnamefont {H.~X.}\
  \bibnamefont {Tang}},\ }\bibfield  {title} {\bibinfo {title} {Two-colour
  dissipative solitons and breathers in microresonator second-harmonic
  generation},\ }\href@noop {} {\bibfield  {journal} {\bibinfo  {journal} {Nat.
  Commun.}\ }\textbf {\bibinfo {volume} {14}},\ \bibinfo {pages} {2798}
  (\bibinfo {year} {2023})}\BibitemShut {NoStop}%
\bibitem [{\citenamefont {Perego}\ \emph {et~al.}(2021)\citenamefont {Perego},
  \citenamefont {Mussot},\ and\ \citenamefont {Conforti}}]{perego2021theory}%
  \BibitemOpen
  \bibfield  {author} {\bibinfo {author} {\bibfnamefont {A.~M.}\ \bibnamefont
  {Perego}}, \bibinfo {author} {\bibfnamefont {A.}~\bibnamefont {Mussot}},\
  and\ \bibinfo {author} {\bibfnamefont {M.}~\bibnamefont {Conforti}},\
  }\bibfield  {title} {\bibinfo {title} {Theory of filter-induced modulation
  instability in driven passive optical resonators},\ }\href@noop {} {\bibfield
   {journal} {\bibinfo  {journal} {Phys. Rev. A}\ }\textbf {\bibinfo {volume}
  {103}},\ \bibinfo {pages} {013522} (\bibinfo {year} {2021})}\BibitemShut
  {NoStop}%
\bibitem [{\citenamefont {Bessin}\ \emph {et~al.}(2019)\citenamefont {Bessin},
  \citenamefont {Perego}, \citenamefont {Staliunas}, \citenamefont {Turitsyn},
  \citenamefont {Kudlinski}, \citenamefont {Conforti},\ and\ \citenamefont
  {Mussot}}]{Bessin_2019}%
  \BibitemOpen
  \bibfield  {author} {\bibinfo {author} {\bibfnamefont {F.}~\bibnamefont
  {Bessin}}, \bibinfo {author} {\bibfnamefont {A.~M.}\ \bibnamefont {Perego}},
  \bibinfo {author} {\bibfnamefont {K.}~\bibnamefont {Staliunas}}, \bibinfo
  {author} {\bibfnamefont {S.~K.}\ \bibnamefont {Turitsyn}}, \bibinfo {author}
  {\bibfnamefont {A.}~\bibnamefont {Kudlinski}}, \bibinfo {author}
  {\bibfnamefont {M.}~\bibnamefont {Conforti}},\ and\ \bibinfo {author}
  {\bibfnamefont {A.}~\bibnamefont {Mussot}},\ }\bibfield  {title} {\bibinfo
  {title} {Gain-through-filtering enables tuneable frequency comb generation in
  passive optical resonators},\ }\href@noop {} {\bibfield  {journal} {\bibinfo
  {journal} {Nat. Commun.}\ }\textbf {\bibinfo {volume} {10}} (\bibinfo {year}
  {2019})}\BibitemShut {NoStop}%
\bibitem [{\citenamefont {Perego}\ \emph {et~al.}(2018)\citenamefont {Perego},
  \citenamefont {Turitsyn},\ and\ \citenamefont {Staliunas}}]{gtl}%
  \BibitemOpen
  \bibfield  {author} {\bibinfo {author} {\bibfnamefont {A.~M.}\ \bibnamefont
  {Perego}}, \bibinfo {author} {\bibfnamefont {S.~K.}\ \bibnamefont
  {Turitsyn}},\ and\ \bibinfo {author} {\bibfnamefont {K.}~\bibnamefont
  {Staliunas}},\ }\bibfield  {title} {\bibinfo {title} {Gain through losses in
  nonlinear optics},\ }\href@noop {} {\bibfield  {journal} {\bibinfo  {journal}
  {Light Sci. Appl.}\ }\textbf {\bibinfo {volume} {7}},\ \bibinfo {pages} {43}
  (\bibinfo {year} {2018})}\BibitemShut {NoStop}%
\bibitem [{\citenamefont {Skryabin}(2020)}]{Skryabin:20}%
  \BibitemOpen
  \bibfield  {author} {\bibinfo {author} {\bibfnamefont {D.~V.}\ \bibnamefont
  {Skryabin}},\ }\bibfield  {title} {\bibinfo {title} {Coupled-mode theory for
  microresonators with quadratic nonlinearity},\ }\href@noop {} {\bibfield
  {journal} {\bibinfo  {journal} {J. Opt. Soc. Am. B}\ }\textbf {\bibinfo
  {volume} {37}},\ \bibinfo {pages} {2604} (\bibinfo {year}
  {2020})}\BibitemShut {NoStop}%
\bibitem [{\citenamefont {Puzyrev}\ \emph {et~al.}(2021)\citenamefont
  {Puzyrev}, \citenamefont {Pankratov}, \citenamefont {Villois},\ and\
  \citenamefont {Skryabin}}]{skr1}%
  \BibitemOpen
  \bibfield  {author} {\bibinfo {author} {\bibfnamefont {D.~N.}\ \bibnamefont
  {Puzyrev}}, \bibinfo {author} {\bibfnamefont {V.~V.}\ \bibnamefont
  {Pankratov}}, \bibinfo {author} {\bibfnamefont {A.}~\bibnamefont {Villois}},\
  and\ \bibinfo {author} {\bibfnamefont {D.~V.}\ \bibnamefont {Skryabin}},\
  }\bibfield  {title} {\bibinfo {title} {Bright-soliton frequency combs and
  dressed states in $\chi^{(2)}$ microresonators},\ }\href@noop {} {\bibfield
  {journal} {\bibinfo  {journal} {Phys. Rev. A}\ }\textbf {\bibinfo {volume}
  {104}},\ \bibinfo {pages} {013520} (\bibinfo {year} {2021})}\BibitemShut
  {NoStop}%
\bibitem [{\citenamefont {Puzyrev}\ and\ \citenamefont
  {Skryabin}(2023)}]{skr2}%
  \BibitemOpen
  \bibfield  {author} {\bibinfo {author} {\bibfnamefont {D.~N.}\ \bibnamefont
  {Puzyrev}}\ and\ \bibinfo {author} {\bibfnamefont {D.~V.}\ \bibnamefont
  {Skryabin}},\ }\bibfield  {title} {\bibinfo {title} {Carrier-resolved
  real-field theory of multi-octave frequency combs},\ }\href@noop {}
  {\bibfield  {journal} {\bibinfo  {journal} {Optica}\ }\textbf {\bibinfo
  {volume} {10}},\ \bibinfo {pages} {770} (\bibinfo {year} {2023})}\BibitemShut
  {NoStop}%
\bibitem [{\citenamefont {Lucia}\ \emph {et~al.}(2014)\citenamefont {Lucia},
  \citenamefont {Huang}, \citenamefont {Corbari}, \citenamefont {Healy},\ and\
  \citenamefont {Sazio}}]{DeLucia:14}%
  \BibitemOpen
  \bibfield  {author} {\bibinfo {author} {\bibfnamefont {F.~D.}\ \bibnamefont
  {Lucia}}, \bibinfo {author} {\bibfnamefont {D.}~\bibnamefont {Huang}},
  \bibinfo {author} {\bibfnamefont {C.}~\bibnamefont {Corbari}}, \bibinfo
  {author} {\bibfnamefont {N.}~\bibnamefont {Healy}},\ and\ \bibinfo {author}
  {\bibfnamefont {P.~J.~A.}\ \bibnamefont {Sazio}},\ }\bibfield  {title}
  {\bibinfo {title} {Optical fiber poling by induction},\ }\href@noop {}
  {\bibfield  {journal} {\bibinfo  {journal} {Opt. Lett.}\ }\textbf {\bibinfo
  {volume} {39}},\ \bibinfo {pages} {6513} (\bibinfo {year}
  {2014})}\BibitemShut {NoStop}%
\bibitem [{\citenamefont {De~Lucia}\ \emph {et~al.}(2019)\citenamefont
  {De~Lucia}, \citenamefont {Bannerman}, \citenamefont {Englebert},
  \citenamefont {Nunez~Velazquez}, \citenamefont {Leo}, \citenamefont {Gates},
  \citenamefont {Gorza}, \citenamefont {Sahu},\ and\ \citenamefont
  {Sazio}}]{de2019single}%
  \BibitemOpen
  \bibfield  {author} {\bibinfo {author} {\bibfnamefont {F.}~\bibnamefont
  {De~Lucia}}, \bibinfo {author} {\bibfnamefont {R.}~\bibnamefont {Bannerman}},
  \bibinfo {author} {\bibfnamefont {N.}~\bibnamefont {Englebert}}, \bibinfo
  {author} {\bibfnamefont {M.~M.~A.}\ \bibnamefont {Nunez~Velazquez}}, \bibinfo
  {author} {\bibfnamefont {F.}~\bibnamefont {Leo}}, \bibinfo {author}
  {\bibfnamefont {J.}~\bibnamefont {Gates}}, \bibinfo {author} {\bibfnamefont
  {S.-P.}\ \bibnamefont {Gorza}}, \bibinfo {author} {\bibfnamefont
  {J.}~\bibnamefont {Sahu}},\ and\ \bibinfo {author} {\bibfnamefont {P.~J.~A.}\
  \bibnamefont {Sazio}},\ }\bibfield  {title} {\bibinfo {title} {Single is
  better than double: theoretical and experimental comparison between two
  thermal poling configurations of optical fibers},\ }\href@noop {} {\bibfield
  {journal} {\bibinfo  {journal} {Opt. Express}\ }\textbf {\bibinfo {volume}
  {27}},\ \bibinfo {pages} {27761} (\bibinfo {year} {2019})}\BibitemShut
  {NoStop}%
\end{thebibliography}%

\end{document}